\documentclass[aps,twocolumn,superscriptaddress]{revtex4}
\usepackage{amsmath,amssymb}
\usepackage{graphics,graphicx}
\usepackage{dcolumn,bm}
\usepackage{psfrag}
\usepackage{color}
\newcommand{\cu}
{\affiliation{Department of Physics, University of Calcutta,
92 Acharya Prafulla Chandra Road, Kolkata 700009, India.}}

\begin{document}

\title{Active-absorbing phase transition and small world behaviour in Ising model on finite addition type networks in two dimensions}

\author{Pratik Mullick}%
\cu
\author{Parongama Sen}%
\cu

\begin{abstract}

We consider the ordering dynamics of the Ising model on a square lattice where an additional fixed number of bonds
connect any two sites chosen randomly. The total number of shortcuts added is controlled by two parameters $p$ and $\alpha$.
The structural properties of the network are investigated which show that that the small world behaviour is obtained along the line
$\alpha=\frac{\ln (N/2p)}{\ln N}$, which separates regions with ultra small world like behaviour and short ranged lattice like
behaviour.
We obtain a rich phase diagram in the $p-\alpha$ plane showing the existence of different types of active and absorbing states
to which Ising model evolves
and their boundaries.
\end{abstract}

\maketitle



\section{Introduction}

When an Ising ferromagnet is quenched from an initial random configuration corresponding to a very
high temperature $T_i$ to a low temperature $T_f$ (below the critical temperature $T_C$), it
undergoes ordering process and gives rise to the formation of domains \cite{brayc2}.
Ising model in one dimension undergoing zero temperature Glauber dynamics always reaches its ground state,
which is the consensus state with all up (down) spins.
However, in higher dimensions the system does not always reach consensus;
in two dimensions sometimes it gets locked in frozen phases.
The two dimensional frozen stable states have higher energy compared to the true ground state but the spins
at the interfaces are unable to flip as that will increase their energy
\cite{lipowski,spirin1,spirin2,barros,olejarz}.
Escape from the frozen states is possible by introduction of a noise or an external field.

Various interesting features appear when the Ising model is studied on a complex network.
As far as static critical behaviour is concerned, one finds a finite temperature phase transition
taking place even when the network is embedded on a one dimensional space.
The critical exponents indicate mean field behaviour.
On the other hand, the dynamical evolution shows the existence of frozen states.
The nature of the final states depends on the type of network and the parameters that defines
the network.

In different types of Watts-Strogatz network (WS), both additional and rewired type, frozen states are seen to
exist for all values of the relevant parameters \cite{herrero,BSen}. 
On an Euclidean network in a one dimensional lattice, connections between neighbours at distance $l$ are added
with probability $P(l) \sim l^{-\alpha}$ \cite{BSen}; here the freezing probability is seen to increase with $\alpha$ and
becomes equal to 1 for large values of $\alpha$.
Addition type WS network on a two dimensional lattice gives rise to frozen states for $p<0.1$ \cite{boyer}
and for the rewired type there exists a critical point $p_c$ above which all the configurations reach
frozen states \cite{herrero}.

When the system reaches either a consensus state or a frozen state, the dynamics no longer
take place and one reaches the so called absorbing state. On the other hand, it may also
happen that the dynamics continue, as for example the three dimensional Ising model \cite{blinkers},
Ising model on Barabasi-Albert network \cite{castellano} and two dimensional Ising model
with longer range interactions \cite{prat2}. Such states are called active states.
Active-absorbing phase transition is a topic of intense research,
the classic example being the directed percolation model \cite{hinri}.

In this paper we consider a network where,
without altering the nearest neighbour connections in the lattice, $B_{ex}$
number of extra bonds to the lattice are added, where
\begin{equation}
B_{ex} = pN^{\alpha}.
\label{bex}
\end{equation}
$N = L \times L$ is the total number of lattice points on the lattice, $p(>0)$ and $\alpha(>0)$ are the two parameters
of the model. Our aim is to find out the fate of the Ising spins at $T=0$ using Glauber dynamics. We are
specifically interested in finding out whether the system remains in an active state or reaches an absorbing state,
the latter is further classified into two types viz. consensus states and frozen states.
We intend to study the nature of the dynamically evolved states for different values of $p$ and $\alpha$,
and show them in the $p-\alpha$ plane along with the phase boundaries.

To obtain any possible correlation with the geometrical properties of the network with the dynamically
evolved phases of the Ising model, we have also investigated the behaviour of the average shortest distance $\langle d \rangle$ and
clustering coefficient $C$ for different values of $p$ and $\alpha$.
The small world behaviour is manifested by the logarithmic variation of the
average shortest distance: $\langle d \rangle \sim \ln N$. 
The average shortest distance $\langle d \rangle$ is the average number of steps required to reach two randomly selected nodes.
It is defined as the mean value of all shortest paths between any two nodes i.e.
\begin{equation}\langle d \rangle = \frac{\sum l_{ij}}{\frac{N(N-1)}{2}},\end{equation} where $l_{ij}$ is the shortest path
between the nodes $i$ and $j$.
Another quantity which is relevant in detecting small world behavior is the clustering coefficient $C$.
Clustering coefficient $c_i$ of node $i$ with degree $k_i$ is given by
\begin{equation}
c_i = \frac{E_i}{\frac{k_i(k_i-1)}{2}},
\end{equation}
where $\frac{k_i(k_i-1)}{2}$ is the number of possible edges between $k_i$ neighbours
and $E_i$ is the number of edges that actually exist.
The average clustering coefficient is given by $C = \frac{1}{N}\sum_ic_i$.
Networks with a small value of $\langle d \rangle$ ($\mathcal{O}(\ln N)$) together with a clustering coefficient much larger than the corresponding
random network are termed as small world network.

\section{Details of the model and quantities calculated}

We consider the Ising spins on the sites of a square lattice
of dimension $L$ undergoing energy minimising single spin flip Glauber dynamics
following a zero temperature quench.
The Hamiltonian for the corresponding interaction between the spins is given by
\begin{equation}
H = -\sum_{\langle ij \rangle}J_{ij}s_is_j,
\end{equation}
where $s_k$ denotes the spin on the $k$-th site, the strength of interaction $J_{ij}=1$
between spins at sites $i$ and $j$.
$\langle ij \rangle$ denotes
summation over connected sites $i$ and $j$.
We use periodic boundary conditions and random asynchronous dynamics.
To add the shortcuts we randomly pick two lattice points which are not the nearest neighbours of
each other and create a bond between them. This step is repeated until the total number of extra connections is $B_{ex}$.
We perform simulations up to a maximum time of $15N$ Monte Carlo steps. The maximum system size simulated
is $L = 64$.

\begin{figure}
\centering
\includegraphics[width=8cm]{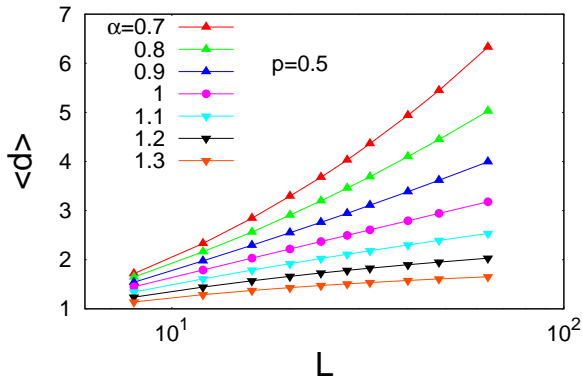}
\caption{Variation of average shortest distance $\langle S \rangle$ with system size $L$ for several values of $\alpha$ keeping
$p = 0.5$.}
\label{svsl}
\end{figure}

\begin{figure}
\centering
\includegraphics[width=8cm]{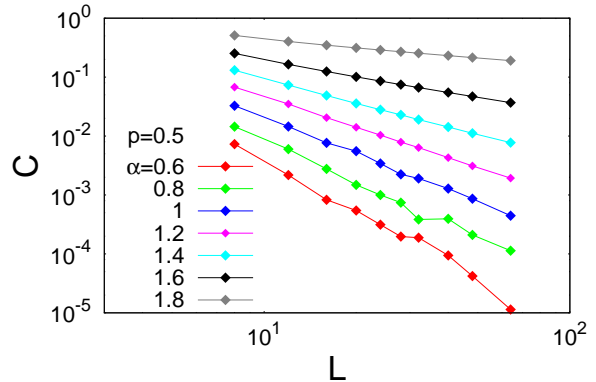}
\caption{Variation of clustering coefficient $C$ with system size $L$ for several values of $\alpha$ keeping
$p = 0.5$.}
\label{clustvsl}
\end{figure}


\begin{figure}
\centering
\includegraphics[width=9.5cm]{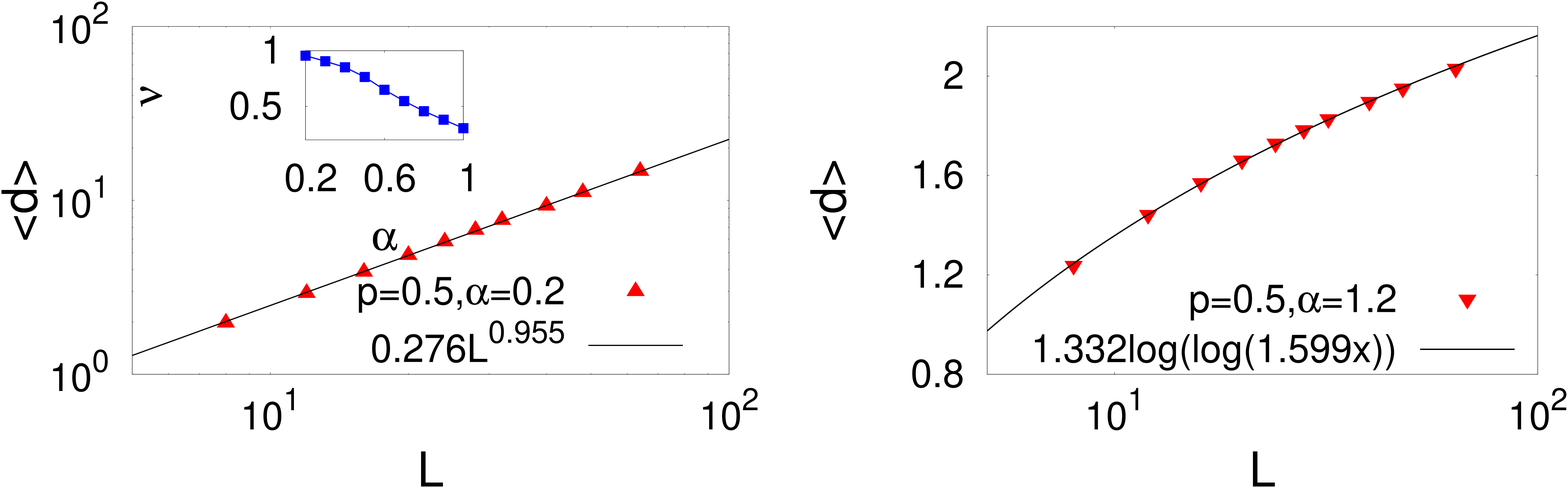}
\caption{Variation of average shortest distance $\langle d \rangle$ with system size $L$ for (a) regular lattice at $p=0.5, \alpha = 0.2$ and for (b) ultra small world network $p=0.5, \alpha = 1.2$. The data clearly shows that (a) $\langle d \rangle \sim L$ for the regular lattice. and (b) $\langle d \rangle \sim \log\log L$ for ultra small world network. Inset of (a) shows the variation of $\nu$ with $\alpha$.}
\label{reglat}
\end{figure}


\begin{figure}
\centering
\includegraphics[width=8cm]{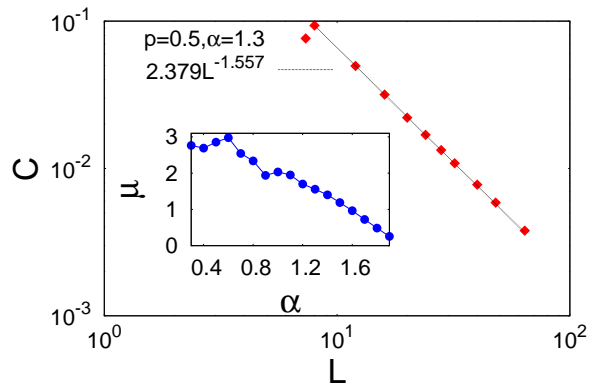}
\caption{Variation of clustering coefficient $C$ with system size $L$ for ultra small world region at $p=0.5, \alpha = 1.3$. The data clearly shows that $C \sim L^{-3/2}$. Inset shows the variation of $\mu$ as a function of $\alpha$.}
\label{clustfit}
\end{figure}

\begin{figure}
\centering
\includegraphics[width=8cm]{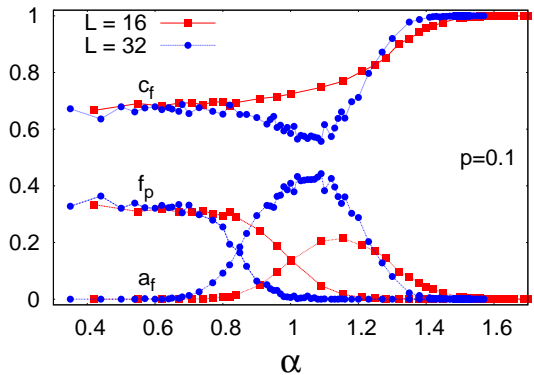}
\caption{Variation of fraction of consensus states $c_f$, freezing probability $f_p$ and fraction of active states $a_f$ as a function
of $\alpha$ keeping $p=0.1$ for two different system sizes.}
\label{freez}
\end{figure}

\begin{figure}
\centering
\includegraphics[width=8cm]{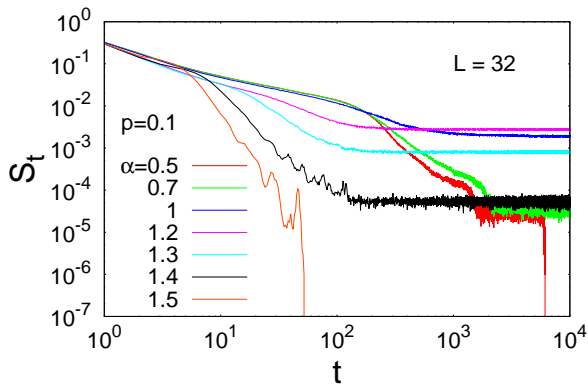}
\caption{Variation of fraction of spin flips $S_t$ as function of time $t$ for several values of $\alpha$ keeping $p=0.1$ for a
$32 \times 32$ system.}
\label{spinflips}
\end{figure}


We first calculate the small world features of the network.
To characterise the small world behaviour we calculate average shortest distance $\langle d \rangle$ and
clustering coefficient $C$ for several values of $\alpha$ keeping $p$ fixed.
These studies were performed by taking average over 100 different initial network configurations.

Ising model in two dimensions undergoing zero temperature Glauber dynamics can give rise to three types of final states viz.
consensus states, frozen states and active states. We therefore, as a function of $\alpha$, measure fraction of
consensus states $c_f$, fraction of frozen states or the freezing probability $f_p$ and fraction of active states $a_f$.
Consensus states are easily detected by value of magnetisation being equal to $\pm 1$.
To differentiate between frozen states and active states, we calculate the fraction of spin flips $S_t$ as a function
of time; a steady state non-zero value of $S_t$ denotes an active state.
We consider 50 different network configurations and corresponding to each network configuration we consider 100 different initial
spin configurations.

\section{Results}

\subsection{Small world properties of the network}

The variation of average shortest distance $\langle d \rangle$
(Fig. \ref{svsl})
and the clustering coefficient $C$ (Fig. \ref{clustvsl}) are studied as a function of the system size $L$ for several values of $\alpha$ keeping $p=0.5$ fixed.

For $\alpha=1$, we find that $\langle d \rangle \sim \log L$ and $C\sim L^{-2} \sim N^{-1}$, which denotes the small world behaviour.
For $\alpha<1$, we see that $\langle d \rangle \sim L \sim N^{1/2}$ (Fig. \ref{reglat}(a)) and $C$ shows a faster decay with $L$, representing regular lattice like behaviour.
The clustering coefficient is basically zero for the original lattice.
For $\alpha>1$, it is observed that $\langle d \rangle \sim \log\log N$ (Fig. \ref{reglat}(b)) and $C\sim L^{-3/2} \sim N^{-3/4}$ (Fig. \ref{clustfit}), as in an ultra
small world network \cite{cohen}.
Hence for $\alpha<1$, the behaviour is characteristic of a short range lattice which for convenience we refer to
as a regular lattice.

We also attempt a power law fitting for the variation of $\langle d \rangle$ as a function of $L$ as $\langle d \rangle \sim L^{-\nu(\alpha)}$,
and plot the variation of $\nu$ as a function of $\alpha$ (inset of Fig. \ref{reglat}(a)), keeping $p=0.5$ fixed.
Algebraic decay of $C$ vs $L$ curves were also observed ($C\sim L^{-\mu}$), and the exponent $\mu$ was
plotted as a function of $\alpha$ (inset of Fig. \ref{clustfit}), keeping $p=0.5$ fixed.

It is known that if $N/2$ bonds are added randomly to the lattice one gets a small world network \cite{newman}.
For less than $N/2$ number of extra bonds, the system shows characteristics of regular lattice and in the region with
greater than $N/2$ number of extra bonds
the system gradually approaches the behaviour of ultra small world \cite{cohen}.
Since ultimately the number of extra bonds is what actually matters for a fixed $N$,
we expect the equation for the curve corresponding to
small world network with $B_{ex}=N/2$ should be
\begin{equation}
\alpha = \frac{\log (N/2p)}{\log N}.
\label{sweq}
\end{equation}
This line acts as a boundary between regular lattice behaviour and ultra small world behaviour.

\subsection{Dynamically evolved states of the Ising model on the network}

We calculate the fraction of the different types of evolved state as a function of $p$ and $\alpha$ (Fig. \ref{freez}).
It is easy to check the presence of a consensus state (discussed earlier).
The fraction of spin flips $S_t$ (Fig. \ref{spinflips}) was studied to differentiate between two types of
non-consensus states viz. active states and frozen states.
For active states, $S_t$ attains a steady non-zero value with time $t$.
We also measure $\langle S_t \rangle$, the thermal average of the saturation value of $S_t$ and study
its variation with $\alpha$. The data shows a bell shaped peak with the peak occurring at $\alpha \approx 1$ for $p=0.1$.

\begin{figure}
\centering
\includegraphics[width=8cm]{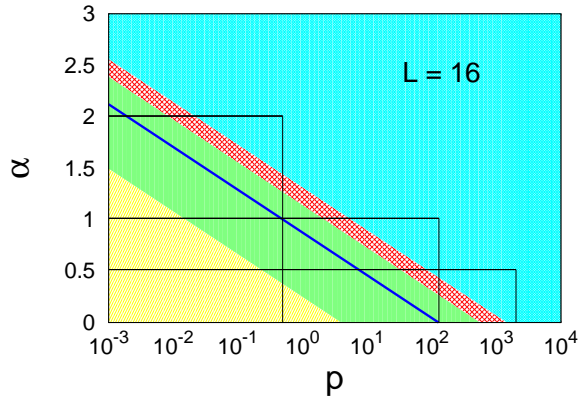}
\caption{$\alpha-p$ phase diagram for $L = 16$. The yellow region: (a) coexistence of consensus states and frozen states; green region: (b) coexistence of consensus, frozen and active states; red region: (c) consensus states and active states, blue region: (d) consensus states only. The blue line indicates small world network, below the blue line: regular lattice like behaviour, above the blue line: ultra small world network.} 
\label{phase16}
\end{figure}

\begin{figure}
\centering
\includegraphics[width=8cm]{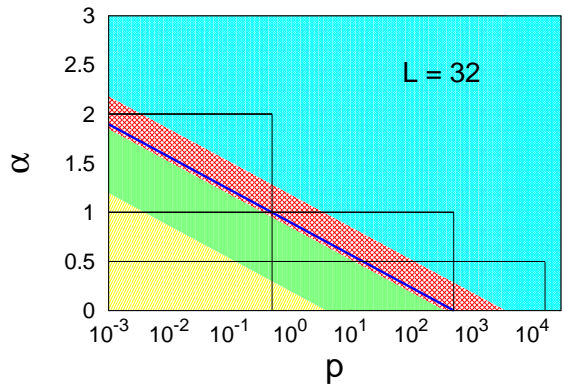}
\caption{Same as Fig. \ref{phase16} with $L = 32$.}
\label{phase32}
\end{figure}

For low values of $p$ and $\alpha$ the system shows regular two-dimensional lattice like behaviour
$f_p\approx 0.33$, $c_f = 1 - f_p$ and $a_f = 0$. Keeping $p$ fixed, when we increase $\alpha$, four distinct
regions appear:
(a) coexistence of consensus states and frozen states ($a_f = 0$)
(b) coexistence of consensus states, frozen states and active states
(c) coexistence of consensus states and active states ($f_p = 0$)
(d) consensus states only ($f_p = 0$, $a_f = 0$).
The boundaries between these regions for a particular $N$ should be marked by particular values of $B_{ex}$, say $b$,
such that the variations are in the form
\begin{equation}
\alpha = \frac{\log (b/p)}{\log N}.
\label{bound}
\end{equation}
Therefore in the $\log$-linear plot of the $p-\alpha$ phase space the boundary curves will be straight lines with a negative
slope. However $b$ can be dependent on the system size $N$.
For a fixed $N$ it is sufficient to detect the boundaries from the results of a single value of $p$ and different values of $\alpha$
as presented is Figs. \ref{phase16} and \ref{phase32}. The results, as expected shows finite size dependence.

\begin{figure}
\centering
\includegraphics[width=8cm]{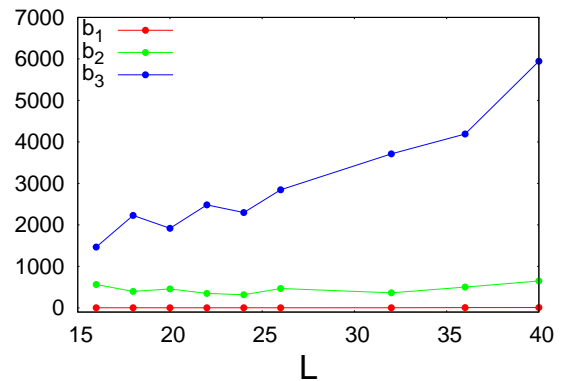}
\caption{Variation of $b_1$, $b_2$ and $b_3$ with system size $L$.}
\label{boundf}
\end{figure}


\section{Discussions and Conclusions}

The maximum number of bonds that can be present in a lattice of total $N$ points is $\sim N^2/2$.
Hence for realistic values of $p$ and $\alpha$, one must have
\begin{equation}
pN^{\alpha}\leq \frac{N^2}{2}.
\label{max}
\end{equation}
For a particular system size, one can determine the maximum value of $p=p_{max}(L,\alpha)$ for a fixed $\alpha$ and the realistic values should be inside the rectangles with sides $\alpha$ and $p_{max}(L,\alpha)$.
We note that such rectangles allow all possible states belonging to the four regions occurring in the phase diagram
for large values of $p$ and $\alpha$.

The yellow region in the phase diagrams corresponds to regular two-dimensional lattice like behaviour consisting
of consensus and frozen states. As the number extra bonds is increased, active states are seen to appear in the
system (green region), a behaviour corresponding to the higher dimensional lattice. Upon further increasing the
number, the fraction of active states ($a_f$) increases. Eventually the system reaches states with fraction of
consensus states $c_f=1$, a characteristic behaviour of fully connected networks.
So this single model encapsulates the behaviour of lattices for dimension $d \geq 2$ by varying the values of the
parameters appropriately.

We have studied the role of $p$ and $\alpha$ for small lattice sizes. Finite size effect will show up in the
phase boundaries since $\log N$ occurs in the denomination of Eq. (\ref{bound}). However, the variation of $b$ with respect to $N$ will also play an important role
and one can study the critical values of $B_{ex}$ which mark the boundaries between the different regions.
We denote these critical values by $b_1$ (separating region (a) and (b)), $b_2$ (separating region (b) and (c))
and $b_3$ (separating region (c) and (d)). We studied $b_1$, $b_2$ and $b_3$ as a function of $L$ as shown in
Fig. \ref{boundf}. It is interesting to note that $b_1$ and $b_2$ have negligible dependence on system size while $b_3$
shows an increase. This indicates that for very large lattices the region with purely consensus
states will be practically non-existent for small $p$ values as observed in \cite{boyer}.

We have studied geometrical properties of the Ising lattice on network for a single value of $p$,
where the behaviour of the system shifts from regular two-dimensional lattice like behaviour to that of higher dimensions
as we increase $\alpha$. This shift is characterised by the values of exponents $\nu$ and $\mu$.
For a different value of $p$ the results will be qualitatively similar, but may show different numerical values depending
on the position of $p$ in the phase diagram.





\end{document}